\newcommand{\ha}{H$\alpha$~}
\newcommand{\nii}{[N~{\sc ii}]$\lambda$~6584\AA~}
\newcommand{\niii}{[N~{\sc ii}]$\lambda$~6548\AA~}
\newcommand{\kms}{{\rm km~s}^{-1}}
\newcommand{\vhel}{V$_{\rm HEL}$}
\newcommand{\arcsec}{$^{\prime\prime}$}
\documentstyle{mn}

\title{Highly supersonic 
motions within the outer features of the $\eta$~Carinae nebulosity}
\author[J. Meaburn, P. Boumis, J.R. Walsh, W. Steffen, 
A.J. Holloway and M. Bryce]{J. Meaburn$^{1}$, P. Boumis$^{1}$, 
J.R. Walsh$^{2}$, W. Steffen$^{1}$,   
A.J.~Holloway$^{1}$, \cr R.J.R~Williams$^{1}$ and M. Bryce$^{1}$ \\
$^{1}$Department of Physics and 
Astronomy, University of Manchester, Oxford
Road, Manchester M13 9PL, UK.\\ 
$^{2}$Space Telescope 
Coordinating Facility, ESO, 
Karl--Schwarzschild--Strasse 2, D-85748 Garching bei
M\"{u}nchen, Germany.}
\date{Accepted ???.
      Received ???.}

\begin{document} 

\maketitle

\begin{abstract}

 Spatially resolved H$\alpha$ and [N{\sc~ii}] line profiles have been
obtained over striking features in the outer regions of the
$\eta$~Carinae nebula. The highly irregular outer `shell' of low
ionization, [N {\sc ii}] bright, knots has been shown to exhibit
radial velocities between $-$1200 and +300 $\kms$ with respect to the
systemic radial velocity, over its perimeter. Furthermore, several
knots have been found which appear to emit only in the \ha line with
radial velocities up to $-$1450~$\kms$.

However, the most intriguing features are a narrow `spike' that
projects through this outer shell and a faint `arc' of emission that
extends well beyond it. The `spike', which exhibits a change of radial
velocity along its length, could be a narrow collimated jet with an
outflow velocity of $\geq$~1000~$\kms$.  In one interpretation the
`arc' is modelled by a conical outflow and mechanisms for generating
[N{\sc~ii}] emission from highly supersonic gas are also discussed.

\end{abstract}

\section{Introduction} The complex nebulosity surrounding
$\eta$~Carinae represents the remnants of ejecta
from this Luminous Blue Variable star (10$^{6.6}$~L$_{\odot}$ 
-- Davidson et al, 1986) during irregular outbursts over the last 300~yrs
(Walborn \& Liller 1977). Investigations
of the kinematics, morphologies 
and ionization mechanisms of the distinct regions identified 
in the sketch in Fig.~1 that comprise
this nebulosity 
have been reported in a series of papers (hereafter referred to as 
Papers 1~--~3 ie. 
Meaburn, Wolstencroft
\& Walsh (1987), 
Meaburn, Walsh \& Wolstencroft (1993~a) and 
Meaburn et al (1993~b) respectively). There is the dusty bi--polar 
Homunculus expanding at $\approx$~650~$\kms$ (Paper 2) surrounded
by a shocked, nitrogen enriched inner shell 
(identified by Walborn et al, 1978 
as S ridge and W arc) with receding,
expansive motions of $\approx$~350~$\kms$ (Paper 1 and
see Dufour, 1989). A knotty
`jet' projects through the NE quadrant of this  
inner shell (Hester et al 1991)
whose episodic, outflow velocity is $\leq$~1500 $\kms$ (Paper~3).

Outside these inner regions a variety of distinctly different
phenomena 
are clearly apparent on UK Schmidt plates, Anglo-Australian 
Telescope (AAT) prime focus plates and the most recent HST images (Fig.~2
and see Sect.~2).
There is the irregular, off--set
outer `shell' of knots partially identified by Walborn et al (1978). 
A feature now marked `spike' in Fig.~1 
(prematurely identified as a jet in 
Paper~1~--~also see fig.~1(c) in Paper 2) projects through this 
irregular outer `shell'.
This can be seen in the HST image in Fig.~2 to be 
$\approx$~0.3$^{\prime\prime}$ wide. 
Also a potentially interesting  feature of the outer nebulosity is 
the faint arc of nebulosity,
marked `arc' in Fig.~1 and outlined by a dashed line. This can be seen on 
a deep AAT prime focus plate to extend
well beyond the perimeter of the outer shell 
in the vicinity of the `spike' and to originate apparently in the vicinity 
of knot E5 (Walborn
et al 1978).

The investigation
of the dramatic kinematics of i) some of the prominent 
knots of the outer `shell', ii) the `spike' (already indicated in
Paper 1) and iii) the `arc' is reported in the present work. 
Spatially resolved profiles of the H$\alpha$ and 
[N{\sc~ii}]$\lambda$6548 and 6584~\AA\ lines have now been obtained
over these regions 
with the EMMI spectrometer on the ESO New Technology Telescope
(NTT) in Chile.

Incidentally, a distance to $\eta$~Carinae of 2.6~kpc will be adopted 
here to be consistent with Papers 2 \& 3. This value is
within the error bars of both 
the measurement of 2.5~kpc by Tapia~et~al (1988)
for the distance to the 
adjacent OB stellar associations 
and the measurement of 2.4~$\pm$0.7~kpc
in Paper 2 based on the expansion velocity and proper motion of
the Homunculus.

\section{Observations and results} 
Longslit spectral observations were obtained on the nights of the 
10 \& 11 March, 1993 using the echelle grating of the EMMI spectrometer
on the NTT but from the remote observing facility at ESO Headquarters 
in Garching. The H$\alpha$ and 
[N{\sc~ii}]$\lambda$ 6548 \& 6584~\AA\
emission lines were isolated by a 90~\AA\ bandwidth interference filter and
the reciprocal dispersion was $\approx$ 5 \AA\ mm$^{-1}$. Cross dispersion
was not employed. Partially resolved line profiles were
obtained from the five slit positions marked 1~--~5 
against a sketch of the outer features of 
$\eta$~Carinae in Fig.~1. Many of these features 
can be seen in the HST image 
in Fig.~2, 
which was obtained with WFPC2 through filter F658N with
an exposure of 400s (two frames coadded). 
In this display the inner features of the 
$\eta$~Carinae nebulosity, that are  
sketched in Fig.~1, are burnt out.
The raw data for this
HST image was obtained from the Hubble Data Archive
and processed using the 1995 set of WFPC2 parameters using WFPC2
pipeline in STSDAS (the images were obtained
in Programme Numbers 5188 and 5239 with W.~Sparks 
and Westphal as the principal investigators respectively). 

A Loral CCD, with 2048~$\times$~2048, 15~$\times$~15~$\mu$m$^{2}$ 
pixels, 
was used as the detector for the spectral observations. With this 
large detector
the H$\alpha$ and [N{\sc~ii}]$\lambda$ 6548 \& 6584~\AA\ lines 
were detected wholly
in the 87th order of the 31.6 grooves mm$^{-1}$ echelle grating
and partially in both the 86th and 88th orders at much lower  
efficiency. However, the inclusion of the 86th and 88th orders
permitted the unambiguous determination of the radial velocities
of particularly high--speed flows.
Spatially resolved 
line profiles were detected in 991 increments, 
each $\equiv$~0.364$^{\prime\prime}$, in the spatial dimension along 
each slit position 
and 1100 channels in the wavelength dimension. The slit width
corresponded to 1.2$^{\prime\prime}$ on the sky ($\equiv$~10~$\kms$). 
The spectra were calibrated in wavelength to $\pm$1~$\kms$
accuracy using the spectra from a Th/Ar emission line lamp. The spectrum  
of a tungsten lamp was used to both flat-field the data and
correct for the blaze of the grating. The integration 
times were 1200 sec.  
The data were processed in the usual way using the FIGARO
reduction software at the Manchester STARLINK node.
 The `seeing' 
was $\leq$ 1$^{\prime\prime}$ throughout these latest spectral
observations.

A first glance at the position-velocity (pv) arrays of profiles 
for Slits~1~--~5 suggested that the radial velocities 
of many of 
the knots of the outer shell
were ambiguous simply because their extents in radial velocity
are near to or greater than 
the separations of the H$\alpha$ and two [N{\sc~ii}]
lines in the spectra. However, as the majority of the high--speed
phenomena are brightest in the [N{\sc~ii}] lines these
ambiguities have been resolved. For example, a feature with a high
approaching radial velocity appears shortwards of
both the 6548\AA\ and 6584\AA\ [N{\sc~ii}] lines by the same amount
and the strengths of the [N{\sc~ii}] lines remain in fixed ratio. 

The negative greyscale representations of the pv  arrays of the
line profiles along Slits~1-5 are shown in Figs.~3(a)-(e)
respectively. These contain all the profiles of 
the H$\alpha$ and the two [N{\sc~ii}] lines 
in the 87th. order as well as high positive radial velocities 
in either the 86th. order (to the
left of the display) or high negative radial velocities in 
the 88th order (to the right).

 The spatial extents of these displays correspond to the 
lengths of the slits marked in Figs.~1~\&~2. The 
various velocity features, which are numbered in the spectra,
are identified along the five slit lengths over enlargements
of the image in Fig.~2 in Figs.~4(a)-(b).
The values of \vhel~of the 
peak of the \nii brightnesses 
for the knots identified in Figs.~4(a)-(b)
are listed in Table 1 along with estimates ($\pm$10 percent)
of the corresponding \nii $/$ H$\alpha$ brightness ratios (again at the 
peak brightnesses for the features have nearly the same extent
in both lines), 
where these ratios are unconfused.

The measurement of \vhel~for a minority of the knots in this way 
is not possible for these are characterised by only a single
feature in the pv arrays in Figs~3(b),~(c)~\&~(d). If these features
are considered real and not instrumental artefacts one
interpretation is that they are produced by knots
with extreme radial velocities where the H$\alpha$
line is dominant. The knots in question are starred in Table~1.
In all cases these appear to be real features in the 
pv arrays for they also are present in the 88th order
to the right hand side of each display.

The feature that poses the greatest problem in this respect
is the looped structure in Fig.~3(d) sub-divided somewhat
arbitarily as separate Slit~4
Knots~1-3. Certainly Knot 3 appears in all three lines 
with \vhel~$\approx$~$-$200~$\kms$ but the interpretation
of the components  identified as Knots~1~\&~2 in 
Fig.~3(d) as part of a coherent feature in the pv array
of \niii profiles, though superficially 
persuasive, must remain doubtful.
If these two knots  have their origin in the \niii profiles then they 
should be clearly present in the three times
brighter \nii profiles. Tests on the data 
have demonstrated their absence in the \nii profiles at the 
appropriate level. This raises
the possibility that Knots~1~\&~2 are exclusively bright  
in the \ha line but with very high negative radial
velocities (starred for Slit 4 in Table~1). Within this
(uncertain) interpretation it is then only a  chance overlap
that creates the impression of a coherent velocity 
feature for Knots~1-3 in Fig.~3(d).
However, such a chance alignment for such very different ranges in radial
velocity seems highly unlikely.
Consequently, in Sect.~3.3, a dual interpretation 
of this strange velocity feature that comprises Slit~4 Knots~1-3
will be presented.

\begin{table}
\centering
\begin{tabular}{||c|c|c|c||}
\hline 
Slit &
Knot &
\vhel &
Ratio \\
N$^{o}$. & N$^{o}.$ & ($\kms$) & \\
\hline
 & 1 & $-$200 & 2.4 \\
 & 2 & $-$50 & - \\
1 & 3 & $-$100 & - \\
 & 4 & +450 & 3.0 \\
 & 5 & +250 & 3.5 \\
\hline
 & 1(a-c) & $-$300, $-$200, $-$100 & - \\
 & 2(a,b) & $-$250, $-$300 & 3.8 \\
 & 2(c,d) & $-$150, $-$50 & 3.8 \\ 
 & 3(a,b) & $-$900, $-$800 & - \\
 & 4 (`spike') & $-$650 & 2.9 \\
2 & 5 & $-$1200 & 2.0 \\
 & 6 & $-$950 & 1.8 \\
 & 7 & +250 & - \\
 & 8 & $-$550 & 2.5 \\
 & 9 & +300 & - \\
 & 10* & $-$1250* & - \\
\hline
 & 1 & $-$250 & 3.5 \\
 & 2 (E5) & $-$140 & 3.5 \\
3 & 3 & $-$200 & 3.5 \\
 & 4 & +250 & - \\
 & 5 (`spike') & $-$850 & 2.2 \\
 & 6* & $-$1450* & - \\
\hline
  & 1* (`arc') & $-$1400* or $-$730 & - \\
4 & 2* (`arc') & $-$1300* or $-$630 & - \\
 & 3 (`arc') & $-$250 & - \\
 & 4 & +300 & - \\
\hline
5 & 1 & $-$400 & - \\
\hline
\end{tabular}
\caption{Motions of the outer shell. (*)~These knots are single features
in the pv arrays. The starred radial velocities assume that they are 
of \ha origin and
alternative unstarred 
values (for Slit~4 Knots~1~\&~2) if of \niii origin. 
The peak intensity ratio of \nii$/$\ha is given in the final
column. Also note that the systemic radial velocity for $\eta$~Carinae
is \vhel~=~$-$7$\kms$}(Paper 2).
\label{table1}
\end{table}

\section{Discussion} 
\noindent New kinematical information has been obtained
for three separate features of the $\eta$~Carinae
nebulosity. The implications for the `outer shell'of knots,
the `spike' and the `arc' will now be discussed
along with consideration of the mechanisms that cause the ionization
of the high-speed ejecta.

\subsection{The `polar blowout' model}

If the knots in the outer `shell' in Figs.~1~\&~2
form part of a coherent feature 
then it is firstly notable that its centroid is
offset by $\approx$~12$^{\prime\prime}$ 
to the NE of $\eta$~Carinae. In any case a 
simple interpretation of this whole feature as a radially expanding
irregular shell is apparently contradicted by the kinematical
observations shown in Figs.~3(a)-(c). This contradiction is epitomised
by the profiles in the pv array along Slits~1,~2~\&~3 in 
Figs.~3(a),(b)~\&~(c) respectively,
which pass over what could be 
the edge of the outer `shell'. In these arrays, certainly detected 
velocity features, which are bright in the [N {\sc ii}] lines, 
extend all the way out to $\approx$~$-$1200~$\kms$ 
(eg. see Slit~2 Knot~5) from 
the systemic radial velocity. 
Single profiles at the systemic radial velocity 
(\vhel~$\approx$~$-$10~$\kms$)
would be expected to characterise the edge 
of a simple, radially expanding shell. Even more extreme \ha bright
knots are suggested out to $-$1450~$\kms$ (eg. see Slit~3 Knot~6 and
Slit~2 Knot~10).

The complex profiles in the pv array in Fig.~3(a) 
from the knots along the north eastern edge of the 
outer shell, covered by Slit~1, exhibit approaching,   
radial  
velocities out to $\approx$~$-$200~$\kms$ from an extended region, 
as well as
a component
at the systemic radial velocity. However, the knots in the 
north western quadrant of the outer `shell', 
covered also by Slit~1,
emit the [N{\sc~ii}] line 
from a region 1$^{\prime\prime}$ across whose profiles have  
FWHMs of $\approx$~80~$\kms$ but are displaced to {\it receding} 
radial velocities of $\approx$~+250~$\kms$ with respect to the
systemic radial velocity. Note that the nearby `lobe' of the inner shell
(identified in Fig.~1) and the 
southern part of this shell (S Ridge in Walborn et al 1978) 
was shown in Paper 1 to emit red-shifted line
profiles out to +400~$\kms$ with respect to the systemic radial
velocity (see Slits 12$^{\prime\prime}$N 
and 4, 8 \& 12$^{\prime\prime}$S in fig.~5 of Paper~1). Some of the 
present (and previous - see Papers 1, 2 \& 3) kinematical observations
are then consistent with the `polar blowout' model of Hester
et al (1991) (and its precursor model in Paper~1 - see fig.~8(b)).
What is loosely described as the outer `shell' of knots in Fig.~1 is 
composed nearly exclusively of ionized material flowing towards 
the observer even around its perimeter. 
These knots (and the `jet' -- see Paper~3)
must all then be a consequence of an
approaching outflow on the nearside of the bi-polar 
axis of the dusty
Homunculus reflection nebula (Paper~2).
The SE lobe of the 
Homunculus is approaching
and the NW receding from the observer along a common axis tilted
at 33$^{\circ}$ to the plane of the sky. Within the 
`polar blowout' model the `inner shell' and `lobe'
in Fig.~1 are then on the far side of the Homunculus and flowing
away from the observer. 

The `spike' and `arc' in Fig.~1 are similarly 
flowing towards the observer but are elongated 
well beyond the SE end of 
the bi-polar axis of the Homunculus. They would clearly have to 
have an extreme configuration
to make them part of a simple `polar blowout' from the centre
of the Homunculus but are likely to be closely
related to this eruptive event.

\subsection{The `spike' as a jet}

\noindent Of particular interest are the line profiles over the `spike'
intercepted by Slit~2 in Fig.~1. 
These are characterised in the pv array in Fig.~3(b) by a bright
[N{\sc~ii}] emission feature, $\approx$~1~$^{\prime\prime}$ 
across that has a FWHM of only 60~$\kms$ but  
displaced from the systemic radial velocity by $\approx$~$-$650~$\kms$
(Slit~2, Knot~4 in Table~1 and Fig.~3(b)).
A faint velocity component can be traced in the same pv array from this
approaching radial velocity right back to the systemic radial velocity. 
The `spike' is also intercepted by Slit~3 ( Knot~5 in Fig.~3(c) and Table~1)
where it now has \vhel~$\approx$~$-$850~$\kms$.

The present kinematical observations of the `spike' along Slit~2 
should be 
interpreted along with those from it but nearer to $\eta$~Carinae
in Paper~1 (EW slit positions 8$^{\prime\prime}$ \& 12$^{\prime\prime}$~S
-- fig.~5 in that paper). The pv arrays across the `spike' 
for these two previous positions
reveal a spatially resolved velocity feature 
that extends continuously out 
to $\approx$~$-$700~$\kms$ from the
systemic radial velocity.  

 The `spike' identified in Fig.~1 and seen in Fig.~4(b)
has a width of a few arcseconds within the `inner shell'
of collisionally ionized gas (see Paper 1) and decreases from
$\approx$~0.35\arcsec~at 18\arcsec~from $\eta$~Carinae (where
it is intercepted by Slit~2) to 0.25\arcsec~at 26\arcsec~(Slit~3). 
An elongated cavity, with thin outflowing walls was suggested
as an explanation of this feature in Paper~1. It was proposed that the 
spike-like appearance arose as the cavity walls were viewed 
tangentially. The present kinematical observations (which are 
further from $\eta$~Carinae than the earlier ones) and the
HST imagery suggest that the `spike' could be a jet ie. a narrow,
collimated outflow exhibiting some form of acceleration or directional
variation
to produce a change of 
\vhel\ from $-$650~$\kms$ to $-$850~$\kms$ between its
interception by Slits~2~\&~3. 

An estimate of the 
outflow velocity of the `spike', if a jet, depends critically
on its orientation with respect to the plane of the sky.
One guide to this angle, $\alpha$, 
could  be the measured orientation for E5 (see Fig.~1). 
This has been derived from the
proper motion of 5$^{\prime\prime}$/century of E5 from $\eta$~Carinae
(Walborn \& Blanco 1989) which, 
for a distance of 2.6~kpc combined with the radial
velocity difference of $\approx$~$-$140~$\kms$ (Table~1) gives
an outflow velocity of 620~$\kms$ tilted at $\alpha$~=~13$^{\circ}$ to 
the sky for this 
group of knots 
E5 marked in Fig.~1. With this same angle a speed
of $\approx$3800~$\kms$ is then predicted by the measured 
value of \vhel~=~$-$850~$\kms$
for the tip of the `spike'. For a constant velocity away from 
$\eta$~Carinae the `spike' would then have originated in an
eruptive event around 1890. This is near the time of major
outbursts listed by Walborn \& Liller (1977).

Perhaps, the orientation of the `spike' can estimated more realistically
if it assumed that its directional variations
{\it in the plane of the sky}, which give an apparent change of 
15$^{o}$ between where 
it is intercepted by Slits 2 and 3, 
are of similar magnitude to directional changes 
perpendicular to the plane of the sky.
This would cause the radial velocity difference 
of $-$200~$\kms$ between these two
positions even if, more 
realistically, no acceleration is assumed. In this case  
$\alpha~\approx$~37$^{o}$ is predicted with a jet speed of
1077~$\kms$ giving an age of $\approx$~360~yr for the tip of the `spike',
which is before records of the eruptions of $\eta$~Carinae are 
available though this speed must be subject to large uncertainties.

Incidentally, the converging sides of the 
`spike' can be explained as part of a jet
structure if the [N~{\sc ii}] emission is from recollimation shocks 
(cf. Paper~3).  A model of the spectrun of Davidson et al. (1986)  
of an [N~{\sc ii}] bright knot 
in the outer shell of the $\eta$~Carinae nebula 
(by John Raymond and reported in Meaburn et al. (1988))
gave a shock velocity of 140~$\kms$. In any case, the shock velocities
cannot be higher than this value for the small 
knots in the outer shell of $\eta$~Carinae to be ionized by
radiative shocks.
Since [N~{\sc ii}] emission requires the shocks to be 
$\leq$~100~$\kms$,  the
aspect ratio of the convergent part of the spike will be roughly
$v_{\rm jet}/(2~v_{\rm shock})$, i.e. $\geq$~5:1 in projection for
a jet speed of 1077~$\kms$
which is easily satisfied by the observed high aspect ratio
of the `spike'.

\subsection{The `arc' as a conical flow}
\noindent Unusual kinematical features also occur along
Slits~4~\&~5 over the ridge marked `arc' in Fig.~1 which 
appears to originate near the knot marked E5. 
Incidentally, a detailed inspection of the HST image in Fig.~2
reveals that both E5 and the northern end of the `arc' are composed of
a conglomeration of emission line knots as small as 0.3$^{\prime\prime}$
across. It is in the profiles from Slit~4 over this `arc' 
that the strange loop (discussed in Sect. 2) 
in the pv arrays occurs (Slit~4 Knots~1~--~3
in Fig.~3(d) and Table~1).
The kinematical features over E5 are themselves
complex (Fig.~3(c)). A `parabola' of emission in the pv array (Slit~3
extended Knot~2), 
15$^{\prime\prime}$
across, extends out to $-$400~$\kms$ from the systemic radial velocity 
though the peak of the emission from E5 can be seen, 
in the contour map of the \nii surface brightnesses shown 
in Fig.~6, to be 
at \vhel~$\approx$~$-$140~$\kms$. 

Assuming the interpretation of the loop as a coherent feature
of \niii emission (though see the reservations expressed in Sect.~2) 
an attempt has been made to reproduce
the approaching `loop' in the pv array in Fig.~3(d) 
from the `arc' sketched in Fig.~1
using the model
shown in Fig.~7(c). Here a conical shell is assumed to have a ratio 
of thickness to radius of 0.15. The ionized gas within this  
shell is given, somewhat arbitarily, a 
flow velocity along its surface of 1500~$\kms$ (similar
to the well-verified speed of the jet in Paper~3).
The full opening angle of the cone from the 
image in Fig.~2 and the AAT photograph is taken to be 
26$^\circ$. The angle between  
the axis of the cone and the plane of the sky is 18$^\circ$. The 
slit is oriented at 30$^\circ$ with respect to the axis of the  
cone to simulate
approximately its orientation to the `arc' in Fig.~1. The predictions
of this model (Fig.~7(b)) convincingly match the contour plot
of the observed  array of profiles in Fig.~7(a) (if these
are all assumed to be of \niii origin). Uniform volume
emissivity of the line emission within the cone would
give rise to a completely closed ellipse in the
predicted pv array in Fig.~7(b). However, in the model, the volume 
emissivity has been reduced in one 
section of the cone until a reasonable match was achieved
to the brightness variations within the observed
loop in the pv array. This self consistent set of key 
parameters (the opening angle, the speed and the angle 
between the cone's axis and the sky) 
is constrained to within ten percent
by the radial velocity range
of $\sim$~$-$130~to $-$900~$\kms$ respectively in the observed
pv array from the `arc'~(Figs.~3(d) and 7(a)). An outflowing cone of 
line emitting knots, perhaps starting in an, as yet unspecified,
manner
from the brighter E5 group of knots, is therefore suggested 
as the explanation of this arc-like feature. 

The absence of the comparable radial velocity components 
in the \nii line profiles makes this interpretation of the `loop'
uncertain. However, one possible way of suppressing the 
comparable \nii emission features (for Slit~4 ~Knots~1~\&~2) 
with approaching radial velocities could be if a localised, but
foreground, receding flow of
H{\sc~ii} gas was present with sufficient density to absorb only
the most negative high velocity \nii features. 
It may be relevant that there is a positive
velocity extension in the \ha pv array over this for Slit~4 
in Fig.~3~d. This mechanism however requires the electron temperature, 
T$_{e}$,
high enough so that the foreground column of H(n=2) has 
$\tau_{H\alpha}~
\gg$~1 but also T$_{e}$ low 
enough that it is fainter than the [N{\sc~ii}] 
itself and since $\tau_{[N II]}~\ll$~1 this may be difficult.

The alternative interpretation of Slit~4~Knots~1~\&~2 as emitting \ha 
only would of course invalidate the model for the `arc' in Fig.~7(c).
Note that a measured decline in the \nii / \ha ratio is shown in Fig.~5 
which suggests that knots with extreme radial velocities (starred in Table~1)
could be undetectable in both the \niii and \nii lines
but still at H$\alpha$. Incidentally Raymond (1991) 
has demonstrated that pure Balmer line emission can be 
produced by non-radiative shocks in partially neutral gas.
A decisive test of these two alternative interpretations of this loop in
the pv array shown in 
Fig.~3(d) awaits observations of the [S{\sc~ii}]$\lambda$6716 
\& 6731\AA\ lines. 
The looped velocity feature could be detected unambiguously 
in such observations.

\subsection{Ionization mechanisms}
\noindent Similar to E5 (Sect.~3.3), 
all of the major groups of low ionization  
knots that constitute the `outer shell' in Fig.~1
have measurable proper motions away from $\eta$~Carinae 
(Walborn \& Blanco~1989). These large proper 
motions 
alone favour their origin in bullets of plasma ejected from
$\eta$~Carinae (during the outburst 150 years ago) 
but now ploughing into slower moving material 
perhaps from previous 
ejections.
In this case, a fundamental 
problem is immediately encountered. 
If the velocities of the bullets relative to the ambient gas 
are $\geq$~1000~$\kms$, as suggested 
for many of the supersonic features revealed in Figs.~3(a)-(e), then 
bow--shock velocities 
of the same magnitude are anticipated which should generate post--shock 
emission at X--ray
wavelengths at their apices 
(which are observed from both the `lobe' and the region of  
the `spike' and E5 in Fig. 1 by Chlebowski et al (1984)). 
Shock velocities must be 
restricted to $\leq$~100~$\kms$ for the post-shock 
emission of [N {\sc ii}] nebular lines that are observed 
here (e.g. Raymond 1979), which would only occur for material 
passing very obliquely through the extreme wings of any bow-shocks
around such high-speed bullets.

A similar difficulty was encountered in interpreting
[N{\sc~ii}] line profiles along the `jet' in
Paper~3 for it is flowing, episodically, 
at a speed of $\leq$~1500~$\kms$ away from
$\eta$~Carinae. 
It was suggested that the narrow, highly blue-shifted [N{\sc~ii}]
profiles from the 
knots in the `jet' originate in slow shocks driven into
dense bullets rather than from the 
post bow-shock regions around the bullets.

Emission of the highly blue-shifted [N{\sc~ii}] profiles 
from behind a bow--shock around a 
high-speed bullet encountering an $\approx$~100~$\kms$ 
slower shell could also be considered to overcome this problem. 
This mechanism could explain the knots which exhibit
a very narrow range of radial velocities, albeit highly
shifted from the systemic, eg. that are seen in the western part  
of the pv array in Fig.~3(a). 
However,
the large range in radial velocities of some of the phenomena
reported here (eg. over the `spike' and `arc' in Fig.~1) 
are not compatible
with post bow-shock emission from bullets and ambient gas which have 
small differential velocities. 

The fine-scale knottiness of all the emission regions seen in Figs. 2,
4(a) and (b) may suggest that the ejecta fragments take the form of
`grape-shot' rather than individual `bullets'.  Cid-Fernandes et al
(1996) have recently shown how individual clumps in the ejecta of
supernovae can be fragmented into `grape-shot' as they pass through a
reverse shock.  These grape-shot clumps then propagate easily through
the shocked ejecta and shocked ambient medium before colliding
directly with the external ambient medium, at velocities little less
than those at which they were driven by the initial explosion.
Similar processes may well occur in outburst ejecta such as those seen
in $\eta$ Car.

If the individual globules within the cloud of condensations are at least
150 times as dense as the surrounding medium, the reverse shocks
driven into 1200 $\kms$ ejecta will be slow enough to radiate [N~{\sc ii}] 
at
velocities close to the ejecta velocity (cf. Slit 2 Knot 5; in contrast 
1500 $\kms$
ejecta such as Knot 10 may still be non-radiative, or at least have
fast enough reverse shocks to prevent [N~{\sc ii}] emission).  Once the
reverse shocks have traversed the clumps, they will be rapidly
dispersed by turbulence eddies, and mixed with gas from the ambient
medium.  The residual momentum of this lower density gas will drive
weak shocks into the surrounding ambient medium.  Succeeding blobs
will encounter this mildly-shocked ambient medium, which would now
have a 100 $\kms$ or so outward velocity and an increased gas density.
Before they overtook the shock driven by the first generation of
blobs, doubly-shocked ambient gas would generate [N{\sc~ii}] emission at
velocities up to 200 $\kms$.  In particular, this may be the explanation
for structures such as Slit 2 Knots 2(a-d).

While Cid-Fernandes et al (1996) have provided initial theoretical
support for the possibility of such mechanisms, much work beyond the
scope of the present paper is required to fully model the origin and
evolution of grape-shot ejecta.

\section{Conclusions}

1) The `outer shell' of low ionization knots is not expanding radially 
for knots can be found with a large range 
of radial velocities on its perimeter (\vhel~=~ $-$1200~$\kms$ to
+300~$\kms$). Even some knots, emitting \ha only, may have
\vhel~=~$-$1450~$\kms$. Some association with a `polar blowout' is 
favoured.

2) Changes in both radial velocity and direction of the `spike'
favour its interpretation as a continuous `jet' with a speed of
1080~$\kms$.

3) The observations of the `arc' remain ambiguous. A coherent loop in
the pv array of \niii profile 
suggests that it could be a narrow cone with an outflow speed
of 1500~$\kms$. The absence of the most negative velocity
features in the corresponding \nii profiles casts doubt 
on this interpretation. Somewhat artificial means can be 
introduced to overcome this problem. Alternatively, the
the loop is composed of \ha profiles at extreme negative
radial velocities and \niii profiles at more modest radial velocities.
In this case a most fortuitious alignment of velocity components
must occur to give the coherent appearance of the loop. Observations
of [S~{\sc~ii}] profiles are needed to resolve the issue.

4) A `grape-shot' model is proposed to explain the knottiness
of much of the emission seen in HST images.

\section{Acknowledgements} 
 
We wish to thank the staff at the New Technology Telescope, La Silla,
and the remote observing facility, Garching 
for their excellent assistance during these observations.

WS and RJRW are grateful to PPARC for receipt of the
Post Doctoral Research Associateships, AJH a Post Graduate Research
Studentship and MB a University of Manchester Research Fellowship.

\section{Legends}

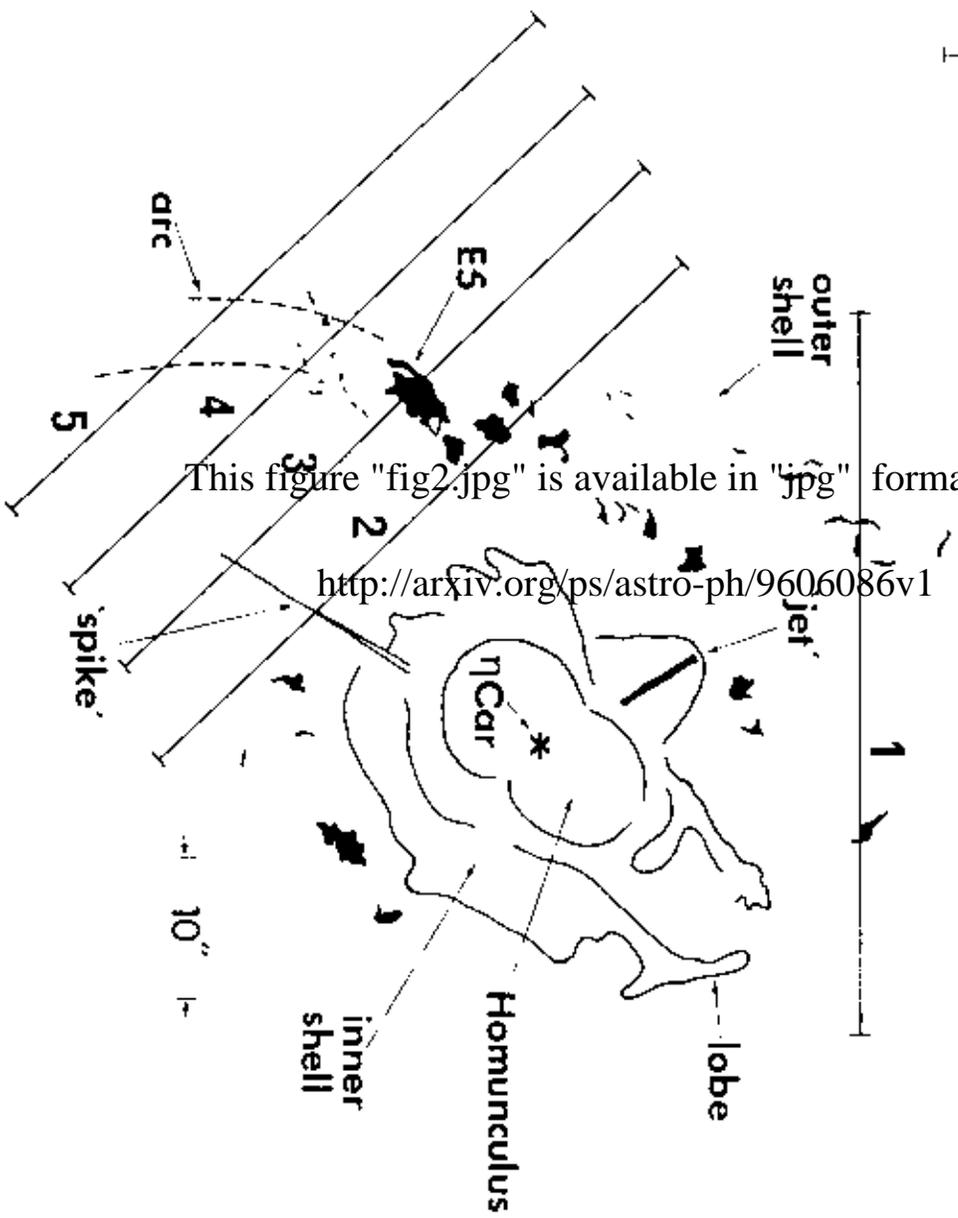
\noindent {\bf Figure 1}

\noindent  The five slit positions, 1-5 are shown against a sketch
of the principal features of the $\eta$~Carinae nebulosity. The
spatial extents of the position--velocity (pv) 
arrays shown in Figs.~3(a)-(e) are indicated by the lengths of the lines
which mark the slit positions.
\vspace{5mm}

\noindent {\bf Figure 2}

\noindent A deep negative print of the HST image of the 
$\eta$~Carinae nebulosity, in the light of 
\nii, in which the features of the outer shell
are revealed. 

\vspace{5mm}

\noindent {\bf Figure 3}

\noindent Negative greyscale representations (whose total extents
are marked in Figs.~1~\&~2) of the pv arrays of  
\ha and [N {\sc ii}] profiles are shown for Slits~1-5
in {\bf (a)}~--{\bf (e)} respectively. The 87th. order
of the echelle spectrum, containing all three lines is in the centre
of the display. Part of the 86th. order is to the left and 
part of the 88th order to the right. The dark vertical
bands are from the background nebular emission
in these three lines. Dark horizontal bands are 
continuous spectra of star images which have also fallen on the slit.

Each separate velocity feature in these displays is identified by
a number which is also marked against the nebular image
in Figs.~4(a)~\&~(b). and listed in Table~1.

\vspace{5mm}

\noindent {\bf Figure 4}

\noindent {\bf (a)} Slit 1 is shown against an enlargement
of the image in Fig.~2. The positions and extents of the velocity
features identified in Figs.~3(a)-(e) are shown. Each is
tentatively called a Knot in Table~1. {\bf (b).} As for (a) but for
Slits~2-5.
\vspace{5mm}

\noindent {\bf Figure 5}

\noindent The \nii$/$H$\alpha$ brightness ratios v. \vhel\ for
the knots on Table~1. All values are for the peaks of the surface
brightnesses of the velocity features in the pv arrays.
\vspace{5mm}

\noindent {\bf Figure 6}

\noindent A contour map of the surface brightnesses of the
pv array  of \nii profiles for 
Knot~2  along Slit~3 (E5 in Fig.~1). The intervals are linear.
\vspace{5mm}

\noindent {\bf Figure 7}

\noindent {\bf (a)} A contour map, with 
linear intervals, of the pv array of [N{\sc~ii}] line profiles 
for slit Pos.~4 which is 
shown in Fig.~3(d). 
The two scales in radial velocity
reflect the uncertain interpretation (as either of \niii or \ha origin)
of Knots~1~\&~2. The `loop' to negative radial velocities   
is from the `arc' marked in Fig.~1. {\bf (b)} The predicted pv array
from the model in (c) which closely matches the observations within
the \niii interpretation of Knots~1 \& 2 in
(a). {\bf (c)} The `arc' in Fig.~1 modelled as a conical outflow 
at 1500 $\kms$ again within the \niii interpretation. 
The slit (the vertical rectangle) 
is at an angle to the axis of this cone to
approximately match its position for the observations.

\bsp


\vspace{5mm}

\begin{thebibliography}{}

\bibitem[\protect\citename{Chlebowski et al}1984]{bu}
Chlebowski, T., Seward, F.D., Swank, J. \& Szymkowiak, A., 1984,
ApJ, 281, 665.

\bibitem[\protect\citename{Cid}1996]{bu}
Cid-Fernandes R., Plewa T., R\'o\.zyczka M., Terlevich R.,
Tenorio-Tagle G., Franco J., Miller W., 1996, MNRAS, in press


\bibitem[\protect\citename{Davidson et al}1986]{bu}
Davidson K., Dufour R.J., Walborn N.R. \& Gull T.R., 1986, ApJ, 305, 867.

\bibitem[\protect\citename{Dufour}1989]{bu}
Dufour R.J., 1989, Rev.Mex.A.A., 18, 87.

\bibitem[\protect\citename{Hester et al}1991]{bu}
Hester J.J., Light R.M., Wetsphal J.A., Currie D.G.,
Groth E.J., Holtzman J.A., Lauer T.R., O'Neil E.J., 1991, AJ, 102, 654.

\bibitem[\protect\citename{Meaburn et al}1987]{bu} 
Meaburn J., Wolstencroft R.D. \& Walsh J.R., 1987, A\&A, 181, 333. (Paper~1)

\bibitem[\protect\citename{Meaburn et al}1988]{bu}
Meaburn J., Wolstencroft R.D., Raymond J.C., Walsh J.R. \& Lopez J.A.,
1988, in `Dust in the Universe', eds. Bailey, M.E. \& Williams, D.A.,
Cambridge University Press, 381.

\bibitem[\protect\citename{Meaburn et al}1993a]{bu}
Meaburn J., Walsh J.R. \& Wolstencroft R.D., 1993a, A\&A, 268, 283. (Paper~2)


\bibitem[\protect\citename{Meaburn et al}1993b]{bu}
Meaburn J., Gehring G., Walsh J.R., Palmer J.W., Lopez J.A., 
Bryce M. \& Raga A.C., 1993b, A\&A, 276, L21. (Paper~3)

\bibitem[\protect\citename{Raymond}1979]{bu} 
Raymond J.C., 1979, ApJ, 39, 1.

\bibitem[\protect\citename{Raymond}1979]{bu}
Raymond J.C., 1991, PASP, 103, 781. 

\bibitem[\protect\citename{Tapia et al}1988]{bu}
Tapia M., Roth M., Marraco H. \& Ruis M.T., 1988, MNRAS, 232, 661.

\bibitem[\protect\citename{Walborn}1977]{bu}
Walborn N.R. \& Liller M.H., 1977, ApJ, 211, 181.

\bibitem[\protect\citename{Walborn et al}1978]{bu}
Walborn N.R., Blanco B.M., Thackeray A.D., 1978, ApJ, 219, 498.

\bibitem[\protect\citename{Walborn}1988]{bu} 
Walborn N.R., Blanco B.M., 1988, PASP, 100, 797.

\end{thebibliography}
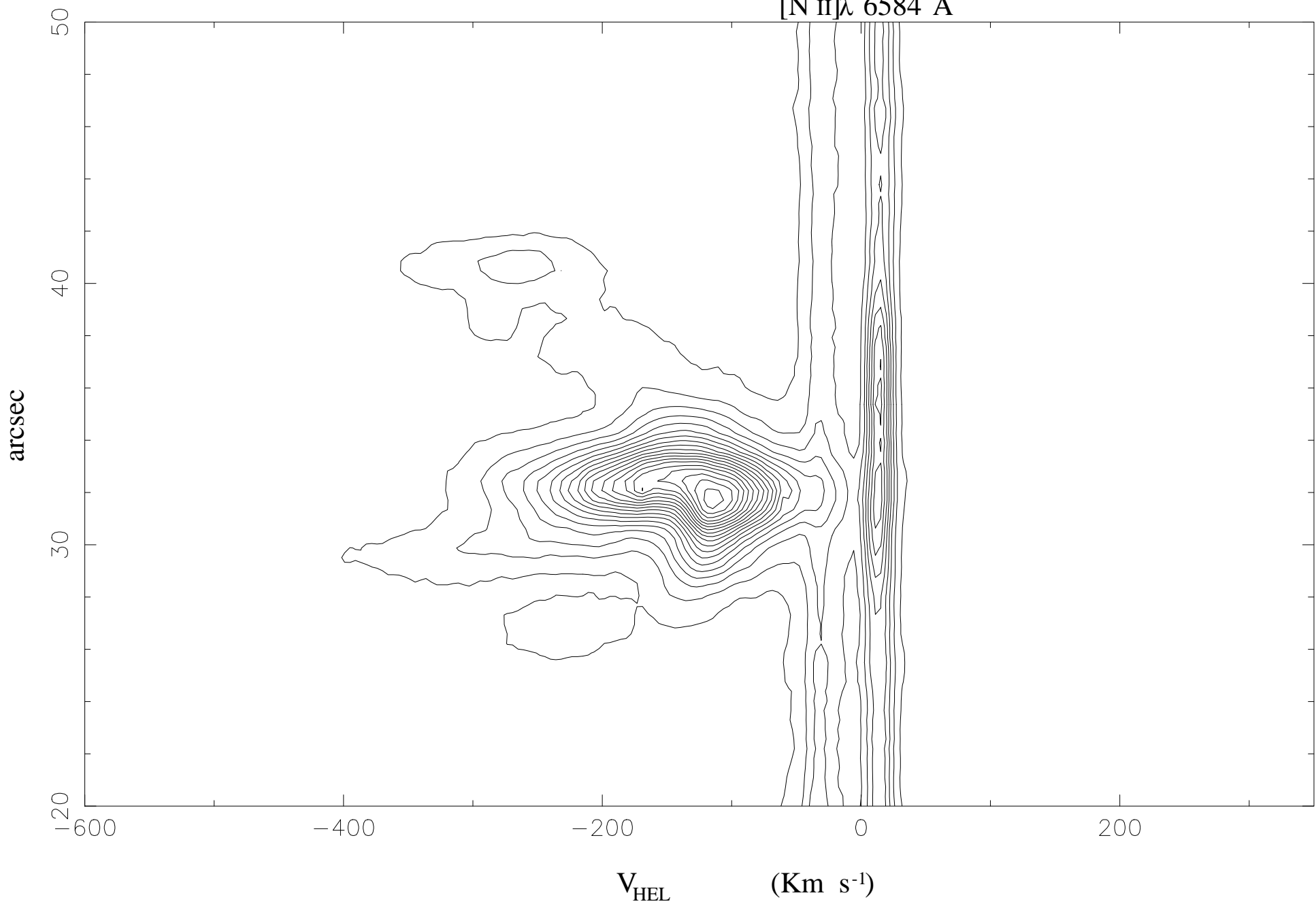
\end{document}